\documentclass[journal=apchd5,manuscript=article]{achemso}
\usepackage{graphicx}
\usepackage{amssymb} 
\usepackage{amsmath} 
\usepackage{float} 

\usepackage[dvipsnames]{xcolor}

\newcommand{\bk}{{\bf k}}
\newcommand{\bE}{{\bf E}}
\newcommand{\br}{{\bf r}}

\title{Inelastic electron-light scattering at dielectric thin films}

\author{Niklas Müller}
\affiliation{Department of Physics, University of Regensburg, 93040 Regensburg, Germany}
\alsoaffiliation{Institute of Physics, University of Oldenburg, 26129 Oldenburg, Germany}
\author{Gerrit Vosse}
\affiliation{Institute of Physics, University of Oldenburg, 26129 Oldenburg, Germany}
\author{Ferdinand Evers}
\affiliation{Department of Physics and Regensburg Center for Ultrafast Nanoscopy, University of Regensburg, 93040 Regensburg, Germany}
\author{Sascha Schäfer}
\affiliation{Department of Physics and Regensburg Center for Ultrafast Nanoscopy, University of Regensburg, 93040 Regensburg, Germany}
\email{sascha.schaefer@ur.de}

\keywords{Ultrafast transmission electron microscopy, photon induced near-field electron microscopy, Fabry-Perot resonances, inelastic electron-light scattering}

\begin{document}
	
\begin{abstract}
In a recently developed methodology termed photon induced near-field electron microscopy (PINEM), the inelastic scattering of electrons off illuminated nanostructures provides direct experimental access to the structure of optical near-field modes and their population. Whereas the inelastic scattering probability can be quantitatively linked to the near field distribution, analytical results for simple light scattering geometries are scarce. Here we derive a fully analytical expression for the coupling strength between free electrons and optical near-fields in planar geometries representing dielectric thin films. Contributions to the overall coupling from the electric field above, below and within the sample are analyzed in detail. By carefully choosing the relative angles between electron beam, light and thin film and by accounting for a broad spectrum of photon energies, we demonstrate that one can imprint optical material properties like the reflectivity onto the electron energy distribution.  
\end{abstract}


\maketitle

\section{Introduction}
Electron microscopy is one of the most versatile tools for studying the structure and properties of materials on atomic length scales, addressing electronic, structural, and spin degrees of freedom \cite{hawkes2019springer}. Recent instrumental and methodological developments have led to the invention of ultrafast transmission electron microscopy (UTEM). UTEM combines the unparalleled spatial resolution of electron microscopy with the high temporal resolution of ultrafast optics \cite{RN77,RN78,RN79,RN2} and allows for the observation of ultrafast structural dynamics and phase transitions \cite{RN80,RN81,RN82,RN83,RN84} as well as magnetic dynamics \cite{PhysRevX.8.031052,berruto2018laser}. More recently, a technique called photon-induced near-field electron microscopy (PINEM) has emerged as a promising tool for the imaging of optical near-fields in the vicinity of illuminated nano-structures \cite{RN54,RN39,RN38,RN86,RN87}. Examples include the spatio-temporally resolved imaging of polariton wave packets in 2D materials \cite{RN88}, the quantum coherent tailoring of the free electron wave function by light fields \cite{RN3,RN36,RN51,RN15,RN52,RN5,RN70} and work towards free-electron/photon correlations in the few photon limit \cite{RN7,RN67,RN65,RN69}. The fundamental mechanism governing PINEM consists of the temporal phase modulation of a free electron wave function within a localized electric field, which results in the inelastic electron scattering off the light field. The process can be interpreted as the stimulated emission or absorption of a photon into or from the near-field triggered by the passing electron and mediated by the broadened momentum spectrum of the near-field. In the past, the abrupt change in the electric field at planar material interfaces was used to imprint an optical phase modulation on an electron beam both in the longitudinal \cite{RN3} and transverse \cite{RN52} direction. An analytical expression for the electron-light coupling strength at a single reflective interface was given in Ref. \cite{RN36}, the more general case of inelastic electron scattering at an illuminated dielectric thin film was not considered analytically, yet.\\


Here, we introduce a fully analytical approach to calculate the interaction strength between a free electron and an optical near-field at a dielectric thin film. The thin film faces form a discontinuity in the spatial distribution of the transversal field component leading to a momentum broadening.  We will show how the overall coupling is influenced by the partial fields before, inside and behind the thin film and how optical resonances can be mapped onto the electron photon coupling. Furthermore{\color{blue},} we give an example how the resonances can be spectrally resolved using chirped light fields.

\section{Description of the model}
We consider a thin film of an optically isotropic material with a thickness $d$ and (complex) refractive index $n$ (Fig.~\ref{fig1}(a,b)). 
As a specific example and without loss of generality, we have chosen silicon as the thin film material in all explicit calculations (dielectric function taken from \cite{RN89,vogt2016development}). 
A fast electron passes through the film with kinetic energy $E_0$, e.g. $E_0=200~$keV at an angle $\beta$.

\subsubsection*{Electric field - Fabry-Perot pattern } 
 
\begin{figure}
 \includegraphics[scale=0.8]{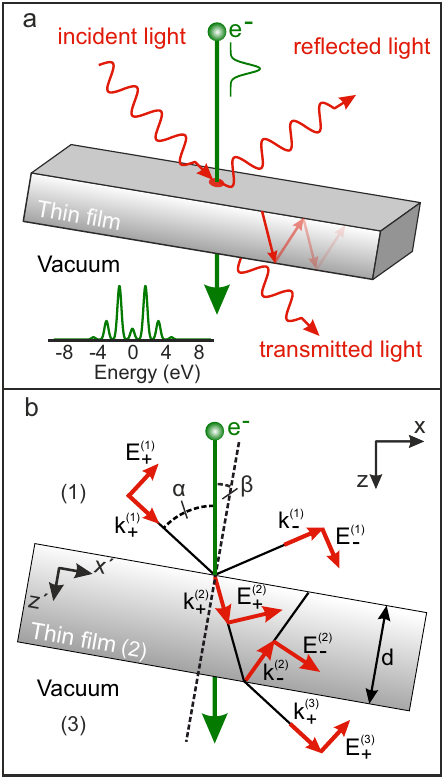}
 \caption{\textbf{Sketch of system geometry} a) A electron pulse passes trough a thin film illuminated by an external light field. 
 The electron absorbs and emits several numbers of photons from the optical near-field of the incident, reflected and transmitted light leading to a modulation of the electron energy distribution with sidebands separated by the photon energy. The light can be reflected multiple times in the thin film forming a Fabry-Perot like resonator for matching light wavelengths. b) Respective wavevectors and electric field vectors for incident, reflected and transmitted light for the case of transverse magnetic (TM) illumination with $\alpha$ being the relative angle between the incident light and the electron direction and $\beta$ being the electron incidents angle relative to the thin film surface normal and $d$ the material thickness.}
\label{fig1}
 \end{figure}

The material is illuminated with an external monochromatic plane-wave light field that constitutes the scattering obstacle for the incoming electron. The incoming light is characterized by an angular frequency $\omega$, a wavevector $\bk$ (incident angle $\alpha$, see Fig.~\ref{fig1}) and a transverse magnetic (TM) polarization. 
At both thin-film interfaces the incident light is partially reflected and transmitted, resulting in interference and the formation of a guided Fabry-Perot mode within the layer.
The resulting Fabry-Perot pattern is controlled by the parameters $d$ and $n$, and on the incident angle of the light wave, $\alpha+\beta$. 

 \subsubsection*{Numerical simulation procedure}
 In each of the regions $i=(1)-(3)$ shown in Fig.~1 (above, inside and below the thin film) the electric field is a superposition of truncated plane waves 
 $\bE=\bE^{(i)}_{\pm} \exp\left( i\left(\bk^{(i)}_{\pm}\br-\omega t \right)\right)$ 
 with wavevectors corresponding to the transmitted and reflected wave components. 
 The individual amplitudes of the plane waves in different regions are linked by the required continuity of the in-plane electric field components (i.e. $E_{x'}$) along the interface, and the out-of-plane (normal) displacement field components ($D_{z'}$); $x'$ and $z'$ denote components in the rotated coordinate system aligned with the thin film axis, Fig.~\ref{fig1}b. 
For calculating the optical field around the thin film, we use a multilayer recursion matrix approach \cite{RN90}, yielding the transmission and reflection coefficients of the in-plane electrical field components (for details, see Supplemental Information 1). 

\begin{figure}
\includegraphics[scale=1]{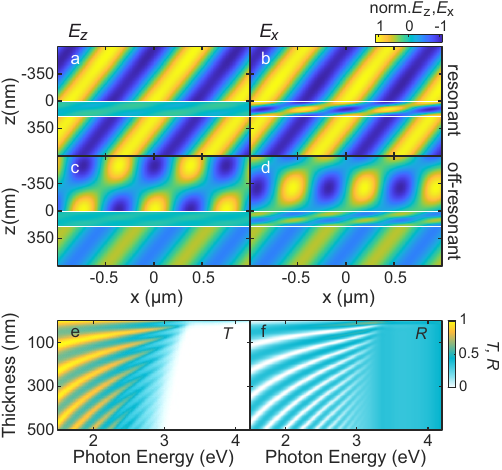}
	\caption{\textbf{Calculated electric field at the sample interface - Fabry-Perot pattern.} a) Exemplary electric field component in z-direction ($\lambda=730~$nm) before, after and in a silicon thin film with a thickness of $d=200~$nm (white region) and an incident angle of $\alpha=57^\circ$. b) Electric field component in x-direction ($\lambda=730~$nm) before, after and in a silicon thin film with a thickness of $d~=~200~$nm (white region) and an incident angle of $\alpha=57^\circ$. c) as a) but for $\lambda=620$. d) as b) but for $\lambda=620$. (Normalized to the respective components of the incident field)  e) Wavelength dependent transmission and f) reflection amplitudes for different silicon film thicknesses.}
	\label{fig2}
\end{figure}
\subsubsection*{Application and ${\bf E}$-field simulation result} 

The electron's incident angle was at first chosen to be normal to the silicon film $\beta=0$ and the incident angle of the light field was fixed at $\alpha=57^\circ$, unless otherwise stated. 
Considering a thin film thickness of $d~=~200~$nm, two exemplary distributions for the $z'$ and $x'$-component of the electric field are plotted in Fig.~\ref{fig2}(a,b) for a case in which Fabry-Perot modes are resonantly excited (illumination wavelength $\lambda=730$~nm) and in Fig.~\ref{fig2}(c,d) for an off-resonant case ($\lambda=620$~nm) respectively. The Fabry-Perot resonances are governed by the relation $2nd\cos\theta^{(2)}=m\lambda$, where $\theta^{(2)}$ is the angle of the light inside the material relative to the surface normal and $m$ is an integer number.

In the resonant case, the transmission is large and the electric field in front of the thin film (region (1)) is dominated by the incident wave which interferes with a much weaker reflected field. Behind the thin film (region (3)) the field is described by a single monochromatic propagating wave phase-shifted with respect to the incident optical field. In the non-resonant case the reflected field in region (1) is much stronger leading to a pronounced checker-board field distribution due to interference with the incident field. 
Below the film (region (3)) the transmitted field is a single propagating plane wave again. Inside the material, off-resonant excitation leads only to weak fields, so that electron-light scattering is mostly determined by the fields outside of the material (region 1 and 3). However, at Fabry-Perot-Resonances enhances field amplitudes can occur inside the material affecting the overall electron-light coupling strength.

The transmittivity and reflectivity, $$T=|E^{(3)}_{+}|^2/|E^{(1)}_{+}|^2, \quad R=|E^{(1)}_{-}|^2/|E^{(1)}_{+}|^2,$$ calculated for a broad spectrum of photon energies and different material thicknesses are plotted in Figure~\ref{fig2}(e,f). Both quantities show a strong dependence on the material thickness as well as the photon energy. For energies larger than the direct band gap of silicon (about 3 eV), the transmission drops significantly due to the onset of lights absorption, whereas for energies smaller than the band gab transmittivity and reflectivity are heavily dominated by the influence of Fabry-Perot resonance modes in the thin film. Specifically, for thickness and wavelength combinations fulfilling the resonance condition, a pronounced decrease in the reflection and an increase in the transmission are observed.

\newcommand{\bA}{{\bf A}}
\newcommand{\bp}{{\bf p}}

\section{Scattering of electrons at modulated ${\bf E}$-fields}
The linear interaction between a moving electron and an optical near-field  can be described semi-classically using the following interaction Hamiltonian in the Coulomb gauge: 
\begin{equation}
	H_I=\frac{e}{m_e}\bA\cdot\hat{\bp}_\text{e},
\end{equation}  
where $e$ is the electron charge, $m_e$ the electron mass, $\bA$ is the vector potential of the optical field and $\hat{\bp}_\text{e}$ the electron momentum operator. 

\subsubsection*{Excerpt of general formalism} 
As previously discussed \cite{RN86,RN39,RN3}, the quantum mechanical state of the electron after leaving the light field can be described by an infinite number of energy ladder states equally spaced by the photon energy, $\left | E_0\pm l \hbar \omega \right >$ with the occupation probabilities given by:
\begin{equation}
	P_l=J_l(2|g|)^2
\end{equation} 
where $l$ is the index of the ladder state and corresponds to the number of net absorbed or emitted photons after the interaction. $J_l$ is the $l$-th order Bessel function of the first kind. The dimensionless coupling constant $g$ is given by
\begin{equation}
	g=\frac{e}{\hbar\omega}\tilde{E}\left(\frac{\omega}{v_e}\right). \label{g}
\end{equation}
where $\tilde{E}(\Delta k)=\int E_z(z) e^{-i\Delta k z}dz$
is the (real-space) Fourier transform of the electric field along the electron trajectory, here taken along the $z$-direction. 
In Eq.~\ref{g} we have $\Delta k=\omega/v_\text{e}$, which equals the momentum transfer experienced by an electron with velocity $v_e$ when it absorbs or emits a photon with frequency $\omega$ from the optical near-field.

\subsubsection*{Application: Scattering off Fabry-Perot pattern} 
\begin{figure}
	\centering
	\includegraphics[scale=0.8]{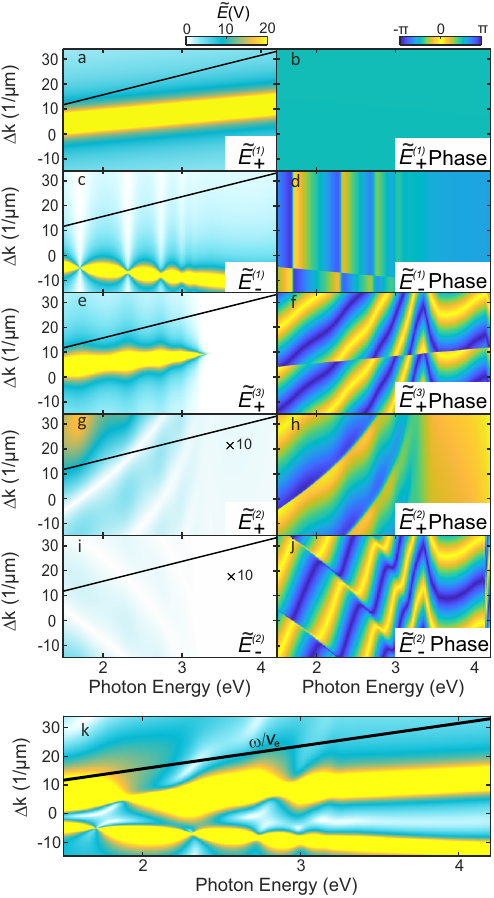}
	\caption{\textbf{Spatial Fourier transforms} of the electric fields (z-component) in three sample regions; incident field chosen as $E^{(1)}_{z,+}= 12\cdot10^6$~V/m. a),b) Amplitude and phase of the Fourier transform of $E^{(1)}_{z,+}$. c),d) $E^{(1)}_{z,-}$. e),f) $E^{(3)}_{z,+}$. g),h) $E^{(2)}_{z,+}$. i),j) $E^{(2)}_{z,-}$. k) Sum of all components. Black lines represent the momentum matching condition $\delta k=\omega/v_e$. 
 }
	\label{fig3}
\end{figure}

In the following, we apply this general formalism to our case. Specifically, the Fourier transform of the optical field at a thin film can be decomposed into additive complex-valued contributions from truncated plane waves in each of the three regions (1) to (3), propagating in the forward ($+$) or backward ($-$) direction (see Supplemental Information 1). Since for electrons in vacuum $\Delta k>k^{(1,3)}_{z,\pm}$, 
we are always in non-phase-matched conditions and the analytical solutions for the Fourier transforms are obtained as:
\begin{subequations}
\begin{align}
&	\tilde{E}^{(1)}_{\pm}(\Delta k)=\pm\frac{iE^{(1)}_{z,\pm}}{\Delta k\mp k^{(1)}_{z,\pm}} \label{E1} \\
&	\tilde{E}^{(2)}_{\pm}(\Delta k)=\pm\frac{iE^{(2)}_{z,+}(e^{-id(\Delta k\mp k^{(2)}_{z,\pm})}-1)}{\Delta k\mp k^{(2)}_{z,\pm}}\\
&\tilde{E}^{(3)}_{+}(\Delta k)=-\frac{iE^{(3)}_{z,+}e^{-id\Delta k}}{\Delta k-k^{(1)}_{z,+}} \label{E3},
\end{align}
\end{subequations}
in which ${E}^{(1)-(3)}_{z,\pm}$ are the z-components in the electron reference frame of the individual field components shown in Fig.~1(b). For the silicon thin film model described above, the individual contributions from each truncated plane wave is shown in Figure~\ref{fig3}(a-j) for a film thickness of $d=200$~nm and photon energies in the visible spectral range and an incident field strength of $E^{(1)}_{z,+}= 12\cdot10^6$~V/m.

The incident field Fourier transform $\tilde{E}^{(1)}_{+}(\Delta k)$ (Fig.~\ref{fig3}a) is peaked along the light line $\Delta k=\omega/c \cos \alpha $. For the incident light truncated at the interface, the light line is broadened,
leading to non-zero components at the position of the electron momentum change condition $\Delta k=\omega/v_e$ (black line). The amplitude of the incident component is independent of the refractive index of the sample and scales with $1/(\Delta k-k_z)$ (Eq.~\ref{E1}) with no phase variation (Fig.~\ref{fig3}b). Equivalently, for the reflected field component $\tilde{E}^{(1)}_{-}(\Delta k)$, maximum Fourier amplitudes are found at the light line of the reflected wave but now exhibiting pronounced variations due to changes in the reflectivity (Fig.~\ref{fig3}c) associated with Fabry-Perot resonances. Overall, the coupling of the reflected wave is generally weaker in our geometry since $\Delta k + k_z$ in the denominator becomes large compared to $\Delta k - k_z$ in the denominator of the incident fields contribution (Eq.~\ref{E1}). Qualitatively this can be understood because the reflected field is counter-propagating with respect to the electron which suppresses an effective coupling. The phase of the reflected fields contribution (Fig.~\ref{fig3}d) is also dominated by the influence of the Fabry-Perot modes with phase jumps of $\pm \pi$ at the resonance positions.

Also the transmitted field contribution $\tilde{E}^{(3)}_{+}(\Delta k)$ (Fig.~\ref{fig3}e) is dependent on the photon energy with the same resonance features as in the reflected field but with a sharp cutoff at the position of the band gap. Furthermore, an additional phase factor $e^{id\Delta k}$ (Eq.~\ref{E3}) is introduced due to the phase shift of the transmitted electrical field compared to the incident field. Hence the phase (Fig.~\ref{fig3}f) has not only a photon energy dependence due to the Fabry-Perot modes but also a pronounced $\Delta k$ dependence.

The contributions from the electric fields inside the thin film (Fig.~\ref{fig3}(g,i)) are smaller by one order of magnitude as compared to the influence of the fields outside the material due to dielectric screening and the normal electron incident ($\beta=0$) considered here. Both Fourier transforms, $\tilde{E}^{(2)}_{\pm}$, are dominated by sinc-like fringes stemming from the rectangular profile of the real space electric fields with corresponding phase jumps (Fig.~\ref{fig3}(h,j)). 
The combination of all contributions, as depicted in Fig.~\ref{fig3}k, is mainly dominated by the contributions from $\tilde{E}^{(1)}_{-}(\Delta k)$, $\tilde{E}^{(1)}_{-}(\Delta k)$ and $\tilde{E}^{(3)}_{+}(\Delta k)$.
\begin{figure}[t]
	\centering
	\includegraphics[scale=0.8]{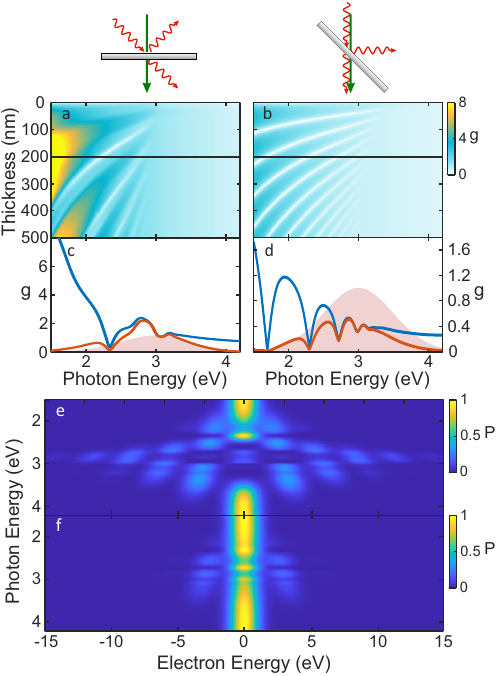}
	\caption{\textbf{Electron light coupling} a) Electron light coupling strength $g$ for different thin film thicknesses and $\alpha=57^\circ$, $\beta=0^\circ$. Corresponding optical geometry is shown in top panel. b) Equivalent to a) but for $\alpha=0^\circ$, $\beta=45^\circ$. c,d)  Line-cut for $g$ for a thickness of $d=200$~nm (blue curves) from a) and b), respectively. Red curve: Coupling strengths multiplied with a Gaussian-shaped spectrum centered around 3 eV (shaded pink area). e) Theoretical PINEM spectrogram as it would be measured when a strongly chirped optical pulse is used with an adjustable delay with respect to the electron pulse. The optical response of the material is directly mapped onto the electron spectrum showing a photon energy dependent decrease and increase of the number of photon sidebands.}
	\label{fig4}
\end{figure}

\section{Electron-light coupling strength} 
 Neglecting the field within the film and expressing the field amplitudes via their analogues in the thin films rotated frame (see Supplemental Information 1) and then replacing them by the reflection and transmission coefficients $r=E^{(1)}_{-}/E^{(1)}_{+}$ and $t=E^{(3)}_{+}/E^{(1)}_{+}$, the overall coupling strength according to Eq.~\ref{g} can be written as:
\begin{align}
\begin{split}
		g=\frac{ie}{\hbar \omega}|E^{(1)}_{+}| & \left(\frac{(1-t\cdot e^{-id \omega/v_e})\sin \alpha}{\omega/v_e-k^{(1)}\cos \alpha} - \right. \\ 
		& \left. \frac{r\sin(\alpha+2\beta)}{\omega/v_e+k^{(1)}\cos(\alpha+2\beta)}\right) . 
\end{split}
	\label{g_complete}
\end{align}
 Eq.~\ref{g_complete}  with $r=-1$, retrieves the expression for the special case of electron-light coupling at a perfect conductor interface derived in Ref.\cite{RN36}.

We calculated the coupling strength $g$ for a variety of thin film thicknesses depicted in Figure~\ref{fig4}a. In comparison to the reflectivity and transmittivity in Figure~\ref{fig2} the overall coupling strength $g$ does not directly resemble the optical features of the thin film. The Fabry-Perot resonances become partially masked by the influence of the $t\cdot e^{id\Delta k}$ interference term in equation~\ref{g_complete} as a result of the phase shift between the fields in front and behind the thin film. 

To increase the visibility of the Fabry-Perot modes the coupling to the reflected field would need to be increased while at the same time the coupling to the incident and transmitted field should be suppressed. This can be achieved by changing the lights incident angle $\alpha$ and the thin films tilt angle $\beta$. Adjusting the angles $\alpha = 0^\circ$ and $\beta = 45^\circ$, the first term in equation~\ref{g_complete} vanishes and the resulting coupling strength, plotted in Figure~\ref{fig4}b, is now solely dependent on the reflectivity $r$ (In fact also the contribution from the field inside the material becomes larger but the overall characteristics are dominated by the reflected field outside the material (see Supplemental Information 2)). This emphasizes the possibility of the PINEM technique to map optical properties directly onto the electron photon coupling. 

\subsection*{Further discussion}
To further elucidate this and to give and example for a possible experimental realization we extract the electron-light coupling strength for a thin film thickness of $d=200$~nm (Fig.~\ref{fig4}c). The coupling between electrons and photons decreases in general with increasing photon energy and has some specific minima and maxima. In an experimental frame a strongly chirped laser pulse could be used together with an ultrashort electron probe pulse to monitor the coupling strength at a specific instantaneous photon frequency by adjusting the relative arrival time of both pulses. Therefore we multiplied the coupling strength with a Gaussian to emulate the temporal pulse form (Fig.~\ref{fig4}d) and calculated the electron energy sideband occupation probability (electron energy distribution after interaction) for every photon energy resulting in a PINEM spectrogram (Fig.~\ref{fig4}e). The photon energy dependent modulation of the coupling strength leads to a decrease and increase of the sideband population if a material resonance is present.

\section{Conclusion}
We derived an analytical expression for the strength of electron-photon scattering at a thin film illuminated by an incident optical field, by using a matching matrix approaches. We calculated the transversal electric field components of the incident and reflected field amplitudes in front, inside and behind the thin film and performing an analytical Fourier transform. The results give insight on how the overall electron-light coupling results from the coupling to the electric fields in the different regions. By altering the incident angles of the electrons and the light field the influence of certain components can be reduced or enhanced. This is useful since certain components carry information about the optical response of the thin film. By choosing optimized angles and applying strongly chirped optical fields, it should be possible to map material depended resonances directly onto the electron sideband population in a PINEM experiment.         

\section{Acknowledgment}
We acknowledge financial support by the DFG within the priority program 1840 "Quantum Dynamics in Tailored Intense Fields", funding by the Volkswagen Foundation as part of the Lichtenberg Professorship "Ultrafast nanoscale dynamics probed by time-resolved electron imaging" and funding by the Free State of Bavaria through the Lighthouse project "Free-electron states as ultrafast probes for qubit dynamics in solid-state platforms" within the Munich Quantum Valley initiative.
	
\section{Author Contributions}
N.M. and S.S. jointly developed the theoretical model. N.M. and S.S. conducted the numerical simulations with the help of G.V.. N.M. and S.S. wrote the manuscript with the help of F.E..

\subsection{Supporting information}
Supporting information: Optical field distribution at a thin film, Choosing the angles $\alpha$ and $\beta$ (PDF)

\bibliography{references}

\end{document}


\section{Optical field distribution at a thin-film}
	In the following, we provide a derivation of the electric fields at an optically isotropic thin-film illuminated with quasi-monochromatic light used to calculate the electron-light coupling constants in the most general case of arbitrary electron beam incidents. 
	For the system sketched in Fig. 1 of the main text, the electric field in each region $i=1-3$ in the coordinate system oriented with the thin film surface normal can be described by a superposition of an incident (+) and reflected (-) plane wave:
	\begin{equation}
		\textbf{E}^{(i)}(\textbf{r}) =\textbf{E}_{+}^{(i)}\exp\left(i \left(k_{z'}^{(i)} z' +k_{x'}^{(i)} x'\right)\right) 
	+\textbf{E}_{-}^{(i)}\exp\left(i \left(-k_{z'}^{(i)} z' +k_{x'}^{(i)} x'\right)\right)
	\end{equation}
	In the case of oblique incidents and TM polarization (magnetic field transversal to the plane of incidents) each field (1)-(3) and (+),(-) can be further decomposed into a component parallel to the interface (along $x'$-direction) and a perpendicular component (along $z'$-direction): $\textbf{E}_{\pm}^{(i)}=E_{x',\pm}^{(i)}\cdot\hat{\textbf{x}}'+E_{z',\pm}^{(i)}\cdot\hat{\textbf{z}}'$, with $\hat{\textbf{x}}',\hat{\textbf{z}}'$ as unit vectors in $x'$ and $z'$ direction in the samples coordinate system (Fig. 1 of the main text). 
	
	The in-plane electric field component needs to be continuous across the interface, resulting in the following matching conditions for the field amplitudes at the interface\cite{RN90}:   
	\begin{equation}
		\begin{bmatrix}
			E^{(i)}_{x',+}\\
			E^{(i)}_{x',-}
		\end{bmatrix} 
		= \frac{1}{t}
		\begin{bmatrix}
			1 & r\\
			r & 1
		\end{bmatrix}
		\begin{bmatrix}
			E^{(i+1)}_{x',+}\\
			E^{(i+1)}_{x',-}
		\end{bmatrix},
	\end{equation}
	where $t$ and $r$ are the Fresnel coefficients for transmission and reflection in the case of TM polarization under oblique incident:
	\begin{equation}
		r=\frac{n\cos(\theta)-\cos(\theta^{(2)})}{n\cos(\theta)+\cos(\theta^{(2)})}
	\end{equation}
	and 
	\begin{equation}
		t=\frac{2\cos(\theta)}{n\cos(\theta)+\cos(\theta^{(2)})}.
	\end{equation}
	with $\theta=\alpha+\beta$ being the incident angle relative to the surface normal outside the thin film and $\theta^{(2)}=\arcsin(\sin(\theta)/n)$ the angle inside the material. Furthermore, for the wave propagation between the first and second interface in the material, a propagation matrix is applied:
	\begin{equation}
		\begin{bmatrix}
			E^{(2)}_{x',+}(z=0)\\
			E^{(2)}_{x',-}(z=0)
		\end{bmatrix} 
		= 
		\begin{bmatrix}
			e^{-i\delta_z'} & 0\\
			0 & e^{i\delta_z'}
		\end{bmatrix}
		\begin{bmatrix}
			E^{(2)}_{x',+}(z'=d)\\
			E^{(2)}_{x',-}(z'=d)
		\end{bmatrix},
	\end{equation}
	with the phase thickness $\delta_z'=\frac{2\pi}{\lambda} n d \cos(\theta^{(2)})$.\\
	Considering the incident field amplitude $E^{(1)}_{x',+}$ and $E^{(3)}_{x',-}=0$, all other field amplitudes can be calculated successively by applying the propagation and matching matrices. The corresponding z-components of the fields are given by: $E_{z',\pm}^{(i)}=E_{x',\pm}^{(i)}\tan(\theta^{(i)})$.
	Afterwards all field amplitudes are converted into the electron reference frame by applying a rotation matrix based on the rotation angle $\beta$ between the electron and the thin film surface normal:
	\begin{equation}
		Rot =
		\begin{bmatrix}
			\cos(\beta)&-\sin(\beta)\\
			\sin(\beta)&\cos(\beta)
		\end{bmatrix} 
	\end{equation}
	
	For the interaction with the electron the longitudinal field component (z-direction) along the electron trajectory (for which we adopt x=0) is important. The different regions fields are multiplied with Heaviside functions $H(z)$ and rectangular functions $\Pi(z)$, respectively, to restrict them to their respective definition regions and to account for the sharp cutoffs at the edges of the film: 
	\begin{subequations}
		\begin{align}
			&E^{(1)}_{z,+}=\left(E_{z,+}^{(1)}\exp(ik^{(1)}_{z,+}z)\right)\cdot(1-H(z))\\ &E^{(1)}_{z,-}=-\left(E_{z,-}^{(1)}\exp(-ik^{(1)}_{z,-}z)\right)\cdot(1-H(z))\\
			&E^{(2)}_{z,+}=\left(E_{z,+}^{(2)}\exp(ik^{(2)}_{z,+}z)\right)\cdot\Pi\left(\frac{z}{d}-\frac{1}{2}\right)\\ &E^{(2)}_{z,-}=-\left(E_{z,-}^{(2)}\exp(-ik^{(2)}_{z,-}z)\right)\cdot\Pi\left(\frac{z}{d}-\frac{1}{2}\right)\\
			&E^{(3)}_{z,+}=\left(E_{z,+}^{(3)}\exp(ik^{(1)}_{z,+}(z-d))\right)\cdot(H(z-d))
		\end{align}
	\end{subequations}
	with $k^{(2)}_{z,\pm}$ being the wavevectors in z-direction inside the material. The wavevectors in z-direction for the incident (+) and reflected (-~) fields are derived from the overall wavevector by using trigonometric realtions and Snells law: $k^{(1)}_{z,+}=k\cos(\alpha)$, $k^{(1)}_{z,-}=k\cos(\alpha+2\beta)$, $k^{(2)}_{z,+}=k\sqrt{1-(\sin(\alpha+\beta)/n)^2}$ and $k^{(2)}_{z,+}=k\sqrt{1-(\sin(\alpha+2\beta)/n)^2}$. For the case of normal electron incidents ($\beta=0$) this simplifies to: $k^{(1)}_{z,+}=k^{(1)}_{z,-}=k^{(1)}_{z}$ and $k^{(2)}_{z,+}=k^{(2)}_{z,-}=k^{(2)}_{z}$. 
	An analytical Fourier transform of Eq. (S7) yields Eq. (4) of the main text. 
	
	\section{Choosing the angles $\alpha$ and $\beta$}
\begin{figure}[H]
		\centering
		\includegraphics[scale=1]{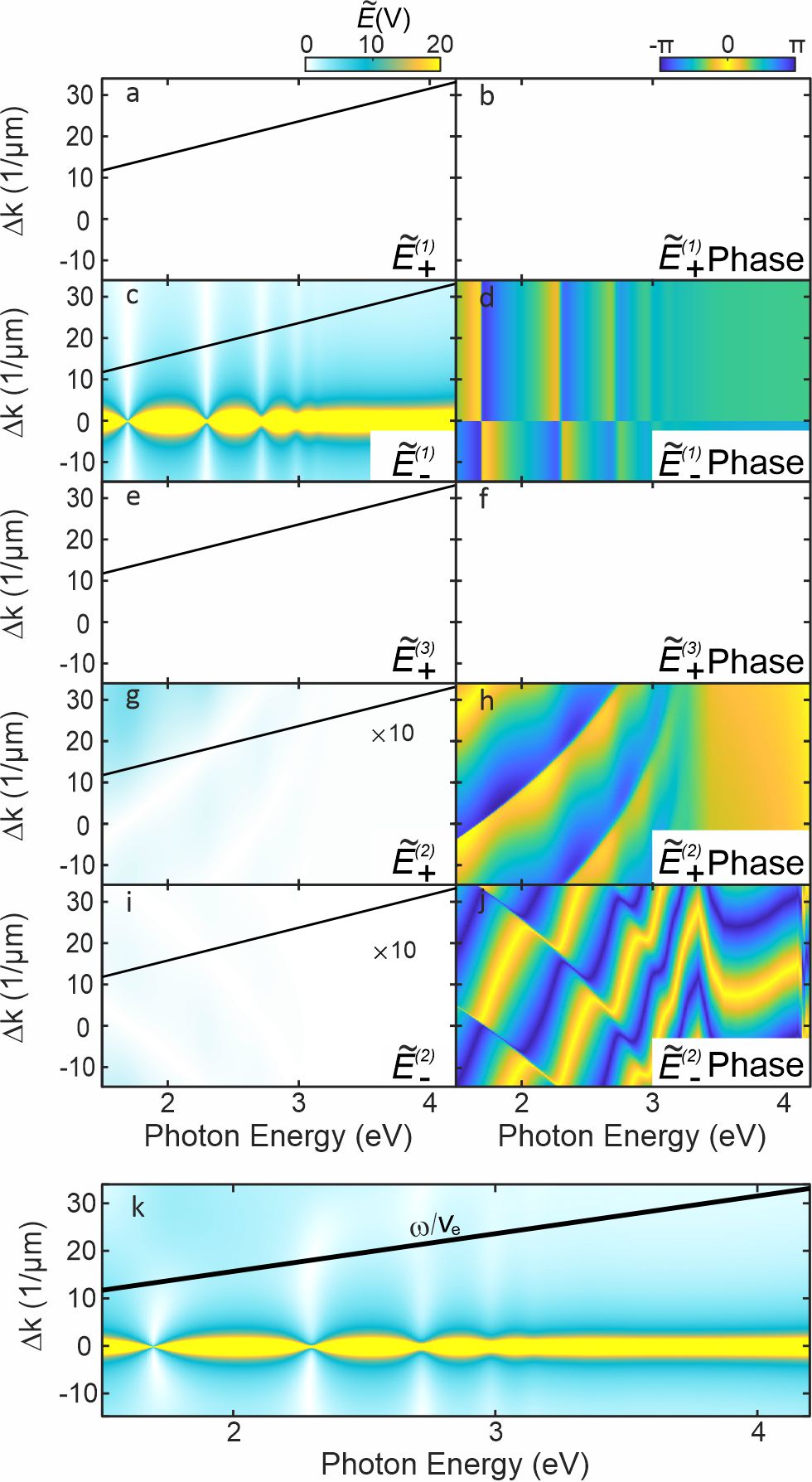}
		\caption{\textbf{Spatial Fourier transform for different field contributions.} Fourier transform of the various field components with an estimated incident field of $E^{(1)}_{z,+}= 12\cdot10^6$~V/m and $\alpha=0^\circ$, $\beta=45^\circ$. a),b) Amplitude and phase of the Fourier transform of $E^{(1)}_{z,+}$. c),d) $E^{(1)}_{z,-}$. e),f) $E^{(3)}_{z,+}$. g),h) $E^{(2)}_{z,+}$. i),j) $E^{(2)}_{z,-}$. k) Sum of all components. The black line represents the momentum matching condition $\delta k=\omega/v_e$. }
		\label{fig3_appendix}
	\end{figure}

	We calculated the electric fields Fourier transforms (Eq. 4 of the main text) for the case of normal electron incident $\beta=0^\circ$ and an angle of $\alpha=57^\circ$ which is the configuration found mostly in transmission electron microscopy where PINEM-like experiments are usually performed. Unfortunately the contribution of the fields inside the material are comparable small and optical resonances of the material are partially masked by the interferences introduced by a relative phase shift between the fields in front and behind the thin film as discussed in the main text. Therefore we have chosen to calculate the fields Fourier transforms in a configuration which suppresses interaction with the incident $E_{z,+}^{(1)}$ and transmitted $E_{z,+}^{(3)}$ field and enhances the coupling to $E_{z,+}^{(2)}$ and $E_{z,-}^{(2)}$, namely $\alpha=0^\circ$ and $\beta=45^\circ$. In this configuration the incident and transmitted light are completely co-propagating with the electrons leading to a zero field component in electron direction, while the reflected light is propagating transversal to the electron beam leading to an electric field purely in electron direction. The resulting Fourier transforms are plotted in Figure \ref{fig3_appendix}.            
	
\bibliography{references}